\documentclass[preprintnumbers,amsmath,amssymb]{revtex4}

\usepackage{dcolumn}
\usepackage{bm}

\begin{document}

\begin{center}
{\Large \textbf{A  Simultaneous  Quantum Secure Direct Communication Scheme between the
Central Party and Other $M$ Parties} $^*$}\\[0.3cm]

GAO Ting$^{1,2,3}$, YAN  Feng-Li$^{3,4}$, WANG Zhi-Xi$^2$\\[0.1cm]

 {\footnotesize $^1$\sl College of Mathematics and Information Science, Hebei Normal University, Shijiazhuang 050016\\
 $^2$ Department of Mathematics, Capital Normal University, Beijing 100037\\
$^3$ CCAST (World Laboratory), P.O. Box 8730, Beijing 100080\\
$^4$ College of Physics Science and Information Engineering, Hebei Normal University, Shijiazhuang 050016}\\[0.3cm]

\begin{minipage} {16cm}
{\noindent\footnotesize \sl We propose a simultaneous quantum  secure direct communication  scheme between one
party and other three parties via four-particle GHZ states and swapping quantum entanglement.  In the scheme,
three spatially separated senders, Alice, Bob and Charlie, transmit their secret messages to a remote receiver
Diana by performing a series local operations on their respective particles according to the quadripartite
stipulation. From  Alice, Bob, Charlie and Diana's Bell measurement results, Diana can infer the secret
messages. If a perfect quantum channel is used, the secret messages are faithfully transmitted from Alice, Bob
and Charlie to Diana via initially shared pairs of four-particle GHZ states without revealing any information to
a potential eavesdropper. As there is no transmission of the qubits
 carrying the secret message  in the public channel, it is
completely secure for the direct secret communication.   This scheme can be considered as a  network of
communication parties where each party wants to communicate secretly with a central
party or server. }\\

\end{minipage}
\end{center}

 One of the most remarkable application of quantum mechanics in quantum information is quantum cryptography, or quantum
key distribution (QKD).   Quantum cryptography exploits the principles  of quantum mechanics to enable provably
secure distribution of private information. QKD  is a protocol that is provable secure, by which private key
bits can be created between two communication parties over a public channel. The key bits can then be used to
implement a classical private key cryptosystem, to enable the parties to correspond securely. QKD can not
prevent eavesdropping, but it can  detect eavesdropping. The basic idea behind QKD is the following fundamental
observation \cite {NC}: Eve can not gain any information from the qubits transmitted from Alice to Bob without
disturbing their states. First of all, by the no-cloning theorem \cite{WZ}, Eve cannot clone Alice's qubit.
Second, another unusual property of quantum mechanics is that, in any attempt to distinguish between two
non-orthogonal quantum states, information gain is only possible at the expense of introducing disturbance to
the signal. These two properties---the quantum no-cloning theorem and the tradeoff between information gain and
disturbance---imply that, by transmitting non-orthogonal qubit states between two parties Alice and Bob, there
is no way to distinguish between non-orthogonal qubit states with certainty. Mainly the first and the most
well-known quantum key distribution protocol BB84 \cite{BB84}, was proposed by   Bennett  and Brassard  in 1984.
As shown in this protocol, Alice and Bob can establish a shared secret key by exchanging single qubits,
physically realized by the polarization of photons, for example. At present, various quantum key distribution
protocols have been proposed, such as Ekert 1991 protocol (Ekert91) \cite {Ekert91}, Bennett-Brassard-Mermin
1992 protocol (BBM92) \cite {BBM92}, B92 protocol \cite {B92} and other protocols \cite {LL, XLG, DLpra68,
DLpra70}.

Different from  QKD whose purpose is to establish a common random key between two parties, a quantum secure
directly communication is to communicate important messages directly without first establishing a random key to
encrypt them.  Shimizu and Imoto \cite {SIpra60, SIpra62} and Beige \emph{et al.} \cite {Beige} proposed novel
quantum secure direct communication (QSDC) schemes, in which the two parties communicate important messages
directly
 and the message is deterministically sent through
the quantum channel, but can be read only after a transmission of an additional classical information for each
qubit. In contrast to the schemes for QKD, which are usually probabilistic, or non-deterministic, the schemes of
quantum secure direct communication are deterministic. Besides, there is an important requirement for direct and
confidential communication that no information is revealed to a potential eavesdropper.  Yan and Zhang \cite
{YZ} put forward  a QSDC scheme using  Einstein-Podolsky-Rosen (EPR) pairs and teleportation  \cite {BBCJPW93}.
In virtue of controlled quantum teleportation \cite {KB} and entangled states we proposed three controlled QSDC
protocols \cite {Gaozna, GYWcp, GYWijmpc}. Entanglement swapping \cite {ZZHE} is a method that enables one to
entangle two quantum systems that do not have direct interaction with each other. Based on entanglement
swapping, three QSDC schemes \cite{GYWnc, GYWjpa2, MZL} are  suggested. All these QSDC schemes \cite{SIpra60,
SIpra62, Beige, YZ, Gaozna, GYWcp, GYWijmpc, GYWnc, GYWjpa2, MZL} are protocols with one communication in the
classical channel and another communication in the quantum channel.

Bostr\"{o}m and Felbinger \cite {BF} presented a communication scheme, the "ping-pong protocol".  It is secure
for key distribution and quasi-secure for direct secret communication if a perfect quantum channel is used.
W\'{o}jcik discussed the security of  the "ping-pong protocol" in a noisy quantum channel  \cite {Wojcik}. Cai
\emph{et al} modified the ping-pong protocol with single photons \cite {CaiLi} and with a similar quasisecurity
property \cite{DLpra69}.
 Deng \emph{et al.}  put forward  two quantum direct
communication protocols, one using Einstein-Podolsky-Rosen pair block \cite {DLL} and another based on polarized
single photons \cite{DLpra69}. A QSDC scheme with quantum superdense coding was also proposed \cite{WDLLLpra71}.
The common character of the QSDC protocols \cite{BF, CaiLi, DLpra69, DLL, WDLLLpra71} is that no additional
classical information is needed for receiver  reading out the secret information.

In this Letter, we generalize the QSDC scheme in Ref.\cite{GYWjpa2} and propose a simultaneous quantum secure
direct communication scheme between the central party and other three parties, which utilizes shared
four-particle Greenberger-Horne-Zeilinger (GHZ) states and entanglement swapping between communicating parties.
  After ensuring the safety of the quantum channel, Alice, Bob and Charlie encode
 secret classical bits by applying predetermined unitary operations on GHZ quartets. The
   secret messages encoded by local operations are faithfully transmitted
from three distant senders (Alice, Bob and Charlie) to a remote receiver (Diana)  without revealing any
information to a potential eavesdropper. The  scheme  can be considered as a  network of communication parties
where each party wants to communicate secretly with a central party or server. Therefore, it is an $M\rightarrow
1$ communication scheme where $M$ can be higher than 3. Indeed,
 it is possible to generalize the scheme  to realize many (more than three) simultaneous
communications among different pairs of parties, such as one for user 1 and server, second for user 2 and
server, third for user 3 and server, $\cdots$, and the $M$-th for user $M$ and server using $(M+1)$-particle GHZ
state $|GHZ\rangle_{M+1}=\frac{1}{\sqrt{2}}(|00\cdots0\rangle+|11\cdots1\rangle)_{12\cdots(M+1)}$ and
entanglement swapping. This scheme, as QKD scheme, can establish a shared key between one party and several
others.

Let us look specially at the case of $M$=3.

(d1) Diana prepares a set of 2$N$ four-particle GHZ states
\begin{equation}\label{2GHZ4}
   |GHZ\rangle=\frac{1}{\sqrt{2}}(|0000\rangle+|1111\rangle),
\end{equation} which are known as  GHZ quadruplets (four-particle GHZ states). Then she shares these entangled
quadruplets of qubits with Alice,
  Bob, and Charlie.
After that they  divide all GHZ quadruplets  into $N$ ordered groups
 $\{\xi(1)_{1234}, \eta(1)_{5678}\}$, $\{\xi(2)_{1234}, \eta(2)_{5678}\}$, $\cdots$, $\{\xi(N)_{1234}$, $\eta(N)_{5678}\}$
  at random. Particles 1 and 5,   2 and 6,  3 and 7, and 4 and 8  of each group belong
to Alice, Bob, Charlie and Diana, respectively.

(d2) Alice and Diana agree on  that Alice can apply the unitary operations
\begin{equation}\label{2operation1}
\begin{array}{cc}
 \sigma_{00}=I=|0\rangle\langle 0|+|1\rangle\langle1|, & \sigma_{01}=\sigma_x=|0\rangle\langle1|+|1\rangle\langle0|, \\
 \sigma_{10}={\rm i}\sigma_y=|0\rangle\langle1|-|1\rangle\langle0|, &
\sigma_{11}=\sigma_z=|0\rangle\langle0|-|1\rangle\langle1|
\end{array}
\end{equation}
 on GHZ quadruplets
$\xi(i)_{1234}$,
 and encode two bits classical information as
\begin{equation}\label{2m12}
\sigma_{00}\rightarrow 00,~ \sigma_{01}\rightarrow 01, ~\sigma_{10}\rightarrow 10, ~\sigma_{11}\rightarrow 11.
\end{equation}
Meanwhile, Diana and Bob, and Diana and Charlie agree on  that Bob and Charlie can only apply unitary operations
\begin{equation}\label{2operation2}
 \sigma_0=I=|0\rangle\langle 0|+|1\rangle\langle1|, ~ \sigma_1=\sigma_x=|0\rangle\langle1|+|1\rangle\langle0|
\end{equation}
 to encode  one bit classical information as following
\begin{equation}\label{2m11}
 \sigma_0\rightarrow 0,~ \sigma_1\rightarrow 1,
\end{equation}
  respectively.

(d3) Alice, Bob, and Charlie  encode their secret messages by applying predetermined unitary operators on
particles 1, 2 and 3, respectively.
 Suppose that Alice, Bob,  Charlie, and Diana initially share GHZ quadruplets $|GHZ\rangle_{1234}$ and
  $|GHZ\rangle_{5678}$,   then the originally entire state of them is
\begin{eqnarray}\label{2original1}
 &&|GHZ\rangle_{1234}\otimes|GHZ\rangle_{5678}\nonumber\\
&=&\frac{1}{4}[\Phi^+_{15}\otimes\Phi^+_{26}\otimes\Phi^+_{37}\otimes\Phi^+_{48}
+\Phi^+_{15}\otimes\Phi^+_{26}\otimes\Phi^-_{37}\otimes\Phi^-_{48}\nonumber\\
&& +\Phi^+_{15}\otimes\Phi^-_{26}\otimes\Phi^+_{37}\otimes\Phi^-_{48}
+\Phi^+_{15}\otimes\Phi^-_{26}\otimes\Phi^-_{37}\otimes\Phi^+_{48}\nonumber\\
&& +\Phi^-_{15}\otimes\Phi^+_{26}\otimes\Phi^+_{37}\otimes\Phi^-_{48}
+\Phi^-_{15}\otimes\Phi^+_{26}\otimes\Phi^-_{37}\otimes\Phi^+_{48}\nonumber\\
&& +\Phi^-_{15}\otimes\Phi^-_{26}\otimes\Phi^+_{37}\otimes\Phi^+_{48}
+\Phi^-_{15}\otimes\Phi^-_{26}\otimes\Phi^-_{37}\otimes\Phi^-_{48}\nonumber\\
&&+\Psi^+_{15}\otimes\Psi^+_{26}\otimes\Psi^+_{37}\otimes\Psi^+_{48}
+\Psi^+_{15}\otimes\Psi^+_{26}\otimes\Psi^-_{37}\otimes\Psi^-_{48}\nonumber\\
&& +\Psi^+_{15}\otimes\Psi^-_{26}\otimes\Psi^+_{37}\otimes\Psi^-_{48}
+\Psi^+_{15}\otimes\Psi^-_{26}\otimes\Psi^-_{37}\otimes\Psi^+_{48}\nonumber\\
&& +\Psi^-_{15}\otimes\Psi^+_{26}\otimes\Psi^+_{37}\otimes\Psi^-_{48}
+\Psi^-_{15}\otimes\Psi^+_{26}\otimes\Psi^-_{37}\otimes\Psi^+_{48}\nonumber\\
&&+\Psi^-_{15}\otimes\Psi^-_{26}\otimes\Psi^+_{37}\otimes\Psi^+_{48}
+\Psi^-_{15}\otimes\Psi^-_{26}\otimes\Psi^-_{37}\otimes\Psi^-_{48}].
\end{eqnarray}
Here
\begin{equation}\label{2EPR}
    \Phi^{\pm}\equiv\frac{1}{\sqrt{2}}(|00\rangle\pm |11\rangle),
    \Psi^{\pm}\equiv\frac{1}{\sqrt{2}}(|01\rangle\pm|10\rangle),
\end{equation}
are  four Bell states (EPR pairs).
 If Alice, Bob, and Charlie want to transmit 01, 0, and 1 to Diana, respectively, then Alice performs
unitary operation $\sigma_{01}$ on particle 1, Bob applies $\sigma_0$ on his particle 2, and Charlie makes
$\sigma_1$ to his particle 3, which imply that the
   state $|GHZ\rangle_{1234}$ is transformed to $\frac{1}{\sqrt{2}}(|0101\rangle+|1010\rangle)_{1234}$.

(d4) Alice, Bob, and Charlie make a Bell measurement on  particles 1 and 5,  2 and 6, and 3 and 7, respectively.
If  Alice obtains measurement result $\Psi^-_{15}$, Bob $\Phi^+_{26}$, and Charlie $\Psi^-_{37}$, then  Diana's
two particles 4 and 8 are in the state $\Phi^+_{48}$ by  the following equation:
\begin{eqnarray}\label{2factual1}
&& \frac{1}{\sqrt{2}}(|0101\rangle+|1010\rangle)_{1234}\otimes|GHZ\rangle_{5678}\nonumber\\
&=&\frac{1}{4}[\Psi^+_{15}\otimes\Phi^+_{26}\otimes\Psi^+_{37}\otimes\Phi^+_{48}
-\Psi^+_{15}\otimes\Phi^+_{26}\otimes\Psi^-_{37}\otimes\Phi^-_{48}\nonumber\\
&& +\Psi^+_{15}\otimes\Phi^-_{26}\otimes\Psi^+_{37}\otimes\Phi^-_{48}
-\Psi^+_{15}\otimes\Phi^-_{26}\otimes\Psi^-_{37}\otimes\Phi^+_{48}\nonumber\\
&& -\Psi^-_{15}\otimes\Phi^+_{26}\otimes\Psi^+_{37}\otimes\Phi^-_{48}
+\Psi^-_{15}\otimes\Phi^+_{26}\otimes\Psi^-_{37}\otimes\Phi^+_{48}\nonumber\\
&& -\Psi^-_{15}\otimes\Phi^-_{26}\otimes\Psi^+_{37}\otimes\Phi^+_{48}
+\Psi^-_{15}\otimes\Phi^-_{26}\otimes\Psi^-_{37}\otimes\Phi^-_{48}\nonumber\\
&&+\Phi^+_{15}\otimes\Psi^+_{26}\otimes\Phi^+_{37}\otimes\Psi^+_{48}
-\Phi^+_{15}\otimes\Psi^+_{26}\otimes\Phi^-_{37}\otimes\Psi^-_{48}\nonumber\\
&& +\Phi^+_{15}\otimes\Psi^-_{26}\otimes\Phi^+_{37}\otimes\Psi^-_{48}
-\Phi^+_{15}\otimes\Psi^-_{26}\otimes\Phi^-_{37}\otimes\Psi^+_{48}\nonumber\\
&&-\Phi^-_{15}\otimes\Psi^+_{26}\otimes\Phi^+_{37}\otimes\Psi^-_{48}
+\Phi^-_{15}\otimes\Psi^+_{26}\otimes\Phi^-_{37}\otimes\Psi^+_{48}\nonumber\\
&&-\Phi^-_{15}\otimes\Psi^-_{26}\otimes\Phi^+_{37}\otimes\Psi^+_{48}
+\Phi^-_{15}\otimes\Psi^-_{26}\otimes\Phi^-_{37}\otimes\Psi^-_{48}].
\end{eqnarray}

(d5) Alice, Bob and Charlie each  tell Diana  that they have measured their respective particles, but  not  the
results of their measurements.

(d6) Diana makes a Bell measurement on her particles 4 and 8 and deduces the outcomes of  Alice, Bob and
Charlie's measurements. By the calculation of entanglement swapping (Eq.(\ref{2original1})) and her measurement
outcome $\Phi^+_{48}$,
 Diana could infer  that the initially whole state $|GHZ\rangle_{1234}\otimes|GHZ\rangle_{5678}$ should collapse to
one of $\Phi^+_{15}\otimes\Phi^+_{26}\otimes\Phi^+_{37}\otimes\Phi^+_{48}$,
$\Phi^+_{15}\otimes\Phi^-_{26}\otimes\Phi^-_{37}\otimes\Phi^+_{48}$,
$\Phi^-_{15}\otimes\Phi^+_{26}\otimes\Phi^-_{37}\otimes\Phi^+_{48}$,
$\Phi^-_{15}\otimes\Phi^-_{26}\otimes\Phi^+_{37}\otimes\Phi^+_{48}$ without Alice, Bob and Charlie's local
operations.

(d7) Diana inquires and gets Alice, Bob and Charlie's measurement results publicly.

(d8) Diana can read out Alice, Bob, and Charlie's secret messages by comparing her calculation result  with
their practical measurement outcomes. From the  measurement results broadcasted by Alice, Bob and Charlie, and
her calculation result, Diana can deduce that Alice, Bob and Charlie, respectively, have  performed unitary
operations $\sigma_{01}$, $\sigma_0$ and $\sigma_1$ on particles 1, 2 and 3,  since by the constraints in (d2)
there is no unitary operation $\sigma_{k_1k_{1'}}\otimes\sigma_{k_2}\otimes\sigma_{k_3}$ ($k_1, k_{1'}, k_2$,
and $k_3 \in \{0, 1\}$) on particles 1, 2, and 3 such that
$\sigma_{k_1k_{1'}}\otimes\sigma_{k_2}\otimes\sigma_{k_3}(\Phi^+_{15}\otimes\Phi^+_{26}\otimes\Phi^+_{37}\otimes\Phi^+_{48})
=\Psi^-_{15}\otimes\Phi^+_{26}\otimes\Psi^-_{37}\otimes\Phi^+_{48}$,
$\sigma_{k_1k_{1'}}\otimes\sigma_{k_2}\otimes\sigma_{k_3}(\Phi^+_{15}\otimes\Phi^-_{26}\otimes\Phi^-_{37}\otimes\Phi^+_{48})
=\Psi^-_{15}\otimes\Phi^+_{26}\otimes\Psi^-_{37}\otimes\Phi^+_{48}$, and
$\sigma_{k_1k_{1'}}\otimes\sigma_{k_2}\otimes\sigma_{k_3}(\Phi^-_{15}\otimes\Phi^-_{26}\otimes\Phi^+_{37}\otimes\Phi^+_{48})=
\Psi^-_{15}\otimes \Phi^+_{26}\otimes\Psi^-_{37}\otimes\Phi^+_{48}$, but only
$\sigma_{01}\otimes\sigma_0\otimes\sigma_1$ satisfying
$\sigma_{01}\otimes\sigma_0\otimes\sigma_1(\Phi^-_{15}\otimes\Phi^+_{26}\otimes\Phi^-_{37}\otimes\Phi^+_{48})
=\Psi^-_{15}\otimes \Phi^+_{26}\otimes\Psi^-_{37}\otimes\Phi^+_{48}$. Therefore  she obtains Alice's message 01,
Bob's 0, and Charlie's 1. Finally, the four spatially separated parties have realized simultaneous deterministic
secure direct communication between one party and other three parties by means of entanglement swapping.

 Similarly, simultaneous  QSDC between
 one party and other $M$ ($M\geq 4$) parties can be realized via shared $(M+1)$-particle GHZ states.

 The crucial point in the proposed scheme is that the qubits carrying the encoded message are not
transmitted in the public channel. Therefore, a potential eavesdropper cannot obtain any information.

 The  above scheme is also a quantum key distribution (QKD) scheme based on four-particle GHZ states
and entanglement swapping. By Alice's measurement result $\Psi^-_{15}$, only both Alice and Diana derive
$\Phi^-_{15} \stackrel{\sigma_{01}}{\longrightarrow} \Psi^-_{15}$, i.e. Alice and Diana obtain
 $\sigma_{01}$ and $\Phi^-_{15}$  in
private. Since Alice's operator $\sigma_{01}$ is certain and her measurement result $\Psi^-_{15}$ is random,
Alice and Diana share two certain bits and two random bits privately. Similarly, Bob and Diana, and Charlie and
Diana share one certain bit  and two random bits  in privacy, respectively. Thus,
 in our proposed scheme, Alice, Bob and Charlie apply one local operation on their respective particles 1, 2
 and 3, Diana  shares two  certain bits and two random bits  with Alice,  one certain bit and two random bits with Bob,
and one certain bit and two random bits with Charlie secretly at the same time.

The security of these protocols is determined by the quality of the quantum channel between the parties. Since
the communication parties  are spatially separated,  after preparation and distribution, the parties may share
an ensemble of noisy GHZ states. In order to share perfect quantum information channel (that is, the shared GHZ
states between the parties are maximally entangled and free of noise), they  first purify noisy GHZ states and
then do some tests  of the security of quantum channel. Suppose that the $(M+1)$ ($M\geq 3$) communication
parties share an ensemble of  identical mixed multi-partite states, they can yield perfect  GHZ states by using
efficient multipartite entanglement distillation protocol--multi-party hashing method \cite{MS} and its
improvement \cite{CL}, and then do the tests. As a matter of fact, as long as the states taking as quantum
information channel are the simultaneous eigenvectors of $(M+1)$ operator products
$S_1=\sigma_x\otimes\sigma_x\otimes\sigma_x\otimes\cdots\otimes \sigma_x$, $S_2=\sigma_z\otimes\sigma_z\otimes
I\otimes\cdots\otimes I$, $S_3=\sigma_z\otimes I\otimes \sigma_z\otimes\cdots\otimes I$, $\cdots$,
$S_{M+1}=\sigma_z\otimes I\otimes I\otimes\cdots\otimes\sigma_z$ with eigenvalue 1, then the quantum channel is
perfect maximally entangled GHZ states, i.e. pure $(M+1)$-particle GHZ states.
 Once the security of the quantum channel is guaranteed,  then no information is leaked to Eve. Therefore
 our proposed protocol is completely secure.

In summary, we present a simultaneous QSDC scheme of three parties and one party by four-particle GHZ states and
entanglement swapping.  In this scheme, the secret messages  are faithfully transmitted from three  senders
Alice, Bob and Charlie  to a  receiver Diana at the same time via initially shared perfect four-particle GHZ
states without revealing any information to a potential eavesdropper.   Since there is no transmission of the
qubit carrying the secret message in the public channel, it is completely secure for direct secret communication
if a perfect quantum channel is used. The scheme can be generalized to more than four parties. Therefore,
simultaneous  QSDC of  $M$ ($M\geq 3$) parties and the central party, who are $(M+1)$ spatially separated
parties, can be realized via initially shared maximally entangled $(M+1)$-particle GHZ states. In this protocol,
quantum storage (quantum memory) is required to store the GHZ states. Certainly, the technique of quantum memory
is not fully developed at present. However, it is a vital ingredient for quantum computation and quantum
information. It is believed that this technique will be available in the future.

{\bf Acknowledgments}

  This work was supported by Hebei Natural Science Foundation of China under
 Grant No. A2004000141 and  No. A2005000140, and  Natural Science Foundation of Hebei Normal University.

\end{document}